\documentclass[prd,a4paper,aps,showpacs,twocolumn,floats,nofootinbib]{revtex4}
\usepackage{epsfig}
\usepackage{graphicx}
\usepackage{color}
\usepackage{amssymb}
\usepackage{amsmath}
\usepackage{hyperref}
\usepackage{url}
\usepackage{natbib}
\usepackage[normalem]{ulem}

\newcommand{\bk}{{\mathbf k}}

\newcommand{\bx}{{\mathbf x}}

\newcommand{\bn}{{\mathbf n}}

\newcommand{\bV}{{\mathbf V}}

\newcommand{\HH}{{\cal H}}

\newcommand{\br}{{\bf r}}
\newcommand{\bnk}{\hat{{\bf k}}}

\newcommand{\al}{\alpha}
\renewcommand{\b}{\beta}
\newcommand{\de}{\delta}
\newcommand{\De}{\Delta}

\newcommand{\ga}{\gamma}

\newcommand{\La}{\Lambda}

\newcommand{\Om}{\Omega}

\newcommand{\si}{\sigma}

\newcommand{\vth}{\vartheta}

\newcommand{\ra}{\rightarrow}

\newcommand{\be}{\begin{equation}}
\newcommand{\ee}{\end{equation}}
\newcommand{\gsim}{\stackrel{>}{\sim}}
\newcommand{\lsim}{\stackrel{<}{\sim}}
\newcommand{\bea}{\begin{eqnarray}}
\newcommand{\eea}{\end{eqnarray}}
\newcommand{\bean}{\begin{eqnarray*}}
\newcommand{\eean}{\end{eqnarray*}}
\newcommand{\dd}{\partial}

\newcommand{\tj}[6]{ \begin{pmatrix}
  #1 & #2 & #3 \\
  #4 & #5 & #6
\end{pmatrix}}
\newcommand*\bra[1]{\left(#1 \right)}
\newcommand*\sbra[1]{\left[#1 \right]}
\newcommand*\bbra[1]{\left\{#1 \right\}}
\newcommand*\abs[1]{\left| #1 \right|}

\newcommand*\braket[1]{\langle#1 \rangle}

\begin{document}

\title{A new method for  the Alcock-Paczy\'nski test}

\author{ Francesco Montanari,  and Ruth Durrer }
\affiliation{D\'epartement de Physique Th\'eorique and Center for Astroparticle Physics, 
Universit\'e de 
Gen\`eve\\ 24 quai Ernest 
Ansermet, CH--1211 Gen\`eve 4, Switzerland}

%\ead{
%\mailto{francesco.montanari@unige.ch},
%\mailto{ruth.durrer@unige.ch}
%}

%\preprint{ }

\date{\today}

\begin{abstract}
We argue that from observations alone, only the transverse power spectrum $C_\ell(z_1,z_2)$ 
and the corresponding correlation function $\xi(\theta,z_1,z_2)$
can be measured and that these contain the full three
dimensional information. We determine the two point galaxy correlation function at linear order in
perturbation theory. Redshift space distortions are taken into account for arbitrary 
angular and redshift separations. We discuss the shape of the longitudinal and the 
transversal correlation functions which are very different from each other and from 
their real space counterpart.
We then go on and suggest how to measure both,  the Hubble parameter, $H(z)$,
and the angular diameter distance, $D_A(z)$, separately from these correlation functions
and perform an Alcock-Paczy\'nski test.

\end{abstract}

\pacs{98.80.-k,98.80.Es}

\maketitle

%-------------------------------------------------------------------------------
\section{Introduction}
Cosmology has become a data driven science. After the amazing success story of the cosmic microwave background (CMB), see~\cite{Komatsu:2010fb,Larson:2010gs,Durrer:2008aa}, which is still ongoing~\cite{planck}, we now also want to profit in an optimal way from actual and future galaxy catalogs. Contrary to the CMB which comes from the two dimensional surface of last scattering, galaxy catalogs are three dimensional and therefore contain potentially more, richer information. On the other hand, galaxy formation is a complicated non-linear process, and it is not clear how much cosmological information about the underlying matter distribution and about gravitational clustering can be gained by
observations of the galaxy distribution. This is the problem of biasing which we do not address in this paper. Here, we simply assume that on large enough scales biasing is linear and local, a hypothesis which might turn out to be too simple~\cite{BeltranJimenez:2010bb}.

When observing galaxies we measure their redshift and angular position. To convert this into a three-dimensional galaxy
catalog we must make an assumption to relate the observed redshift to a distance. For small redshift, the simple
relation $D=z/H_0$ can be used.
Redshift space distortions (RSD) can be taken into account with a
convenient expansion in tripolar spherical harmonics
\cite{Szapudi:2004gh,Papai:2008bd}. This gives an accurate description of the correlation function at small 
scales.
Apart from RSD,  a wrong measurements of $H_0$ will just
rescale the entire catalog but not distort its clustering properties. 

However, if we go out to high redshifts, $z\gtrsim 1$,
non-linear terms in $z$ become relevant, and wrong assumptions about the distance redshift relation can bias the entire catalog.
We therefore believe that it is important to analyze the truly observed catalog, either using the $C_\ell(z_1,z_2)$ spectra introduced in Ref.~\cite{Bonvin:2011bg}, or the angular correlations functions $\xi(\theta,z_1,z_2)$ to describe the observations, and to compare them with their theoretically obtained counterparts.
In this way we truly compare
observations with their theoretical modeling. If, on the other hand, we determine a power spectrum in Fourier space
for the observed catalog, $P(k)$, we have already assumed a cosmology to convert observable redshifts 
into length scales. Therefore, e.g., cosmological parameter estimations using $P(k)$ can at best be viewed as a consistency check.
If the cosmological parameters used to obtain  $P(k)$ agree with those revealed in a Markov Chain--Monte Carlo
parameter estimation, the model is consistent. A thorough error estimation seems however, quite tricky.

We therefore advocate to abandon this `mixed' method in future, high redshift catalogs in favor of the more direct
procedure which compares theoretical models with the directly observed two-point statistics, 
$C_\ell(z_1,z_2)$ and/or $\xi(\theta,z_1,z_2)$.

This paper is structured as follows:
in the next section we explain how to compute  $\xi(\theta,z_1,z_2)$ and its
covariance matrix from the theoretical power spectrum.
In Section~\ref{s:ap} we discuss the longitudinal and transverse correlation functions
and we show how  the baryon acoustic oscillations in
these correlation  functions can be used as an  Alcock-Paczy\'nski 
test~\cite{Alcock:1979mp}.
In Section~\ref{s:con} we conclude.
In Appendix \ref{sec:tripolar} we generalize the expansion in tripolar
spherical harmonics of \cite{Papai:2008bd} to model the RSD of the 
correlation function at arbitrary redshifts.

%-------------------------------------------------------------------------------
\section{Generalities}\label{s:gen}
In an observation which simply counts galaxies,
we measure the galaxy distribution in angular and redshift space.
For simplicity, and since we are interested only in large scales, we assume that its fluctuations are related by a scale independent bias factor $b(z)$ to the matter density fluctuations 
$$\De(\bn,z)=\frac{\rho(\bn,z)-\rho_b(z)}{\rho_b(z)}\,.$$ 
Here $\rho(\bn,z)$ is the true matter density in direction $\bn$ from the observer which we position at $\bx=0$, at measured redshift $z$ and $\rho_b(z)$ is the matter
density of a Friedmann background Universe.

In Ref.~\cite{Bonvin:2011bg}, $\De(\bn,z)$ has been computed including all relativistic effects in first order in perturbation theory. In the present paper we include only the terms which contribute significantly (more than 1\%) under the circumstances discussed here. These are the density fluctuation, the redshift space distortion and, in some cases 
also, the lensing term.
For simplicity, we shall not consider lensing in this work, see however~\cite{Bertacca2012tp}.
Neglecting the other gravitational and Doppler terms, the expression given in 
Ref.~\cite{Bonvin:2011bg} for $\De(\bn,z)$ becomes\footnote{Note that
in \cite{Bonvin:2011bg} $\bn$ denotes the direction of propagation, which is
opposite to the direction of observation considered here.}
\bea
\De(\bn,z) &=& D(\bn,z) - \frac{1}{\HH}\frac{\dd}{\dd r}\left( \bV(r(z)\bn,t(z))\cdot\bn \right) \nonumber \\
&& - \frac{\alpha(z)}{r(z)} \bV(r(z)\bn,t(z))\cdot\bn \;.
\label{e:Delta}
\eea
Here $t(z)$ is conformal time, $D$ is the density fluctuation in comoving gauge,
$\bV$ is the peculiar velocity in longitudinal gauge, and $\HH=aH$ is the
comoving Hubble parameter.
Setting $\al(z)=2/\HH$ corresponds to the sub-leading term 
of the redhsift space distortion, but one can also consider other terms than RSD in the function $\alpha(z)$,
like the boost $-\dot{\HH}/(r\HH^2) \bV\cdot\bn$ calculated in \cite{Bonvin:2011bg},
or a selection function \cite{Raccanelli:2010hk}. For
more details see~\cite{Bonvin:2011bg,Challinor:2011bk} and~\cite{Durrer:2008aa}.
The power spectrum $P(k)$ of density fluctuations (e.g. in synchronous gauge) can be 
calculated with standard Boltzmann solvers~\cite{Lewis:1999bs,Lesgourgues:2011re}.

We define the redshift dependent angular correlation function by
\be\label{e:xi}
\xi(\theta,z_1,z_2) = \langle\De(\bn,z_1) \De(\bn',z_2)  \rangle  ~\mbox{ with }  \bn\cdot\bn' = \cos\theta \,.
\ee
Statistical isotropy implies that $\xi$ depends only on the angle $\theta$ and not on the directions $\bn$ and $\bn'$ separately. In Ref.~\cite{Bonvin:2011bg}, this correlation function is expanded in spherical harmonics,
\be\label{e:cl}
\xi(\theta,z_1,z_2) = \frac{1}{4\pi}\sum_\ell (2\ell+1)C_\ell(z_1,z_2)P_\ell(\cos(\theta)) 
\ee
where $P_\ell(\mu)$ are the Legendre polynomials and we call $C_\ell(z_1,z_2)$ the angular power spectrum.
In principle one can now go on and expand the $z_1$ and $z_2$ dependence in terms of orthonormal functions.
This direction has been explored in Ref.~\cite{Rassat:2011aa}. Since the background Universe and 
therefore also the correlations depend on  $z_1$ and $z_2$ separately and not simply on $r(z_1)-r(z_2)$, 
we refrain from this step here. Clearly, $\xi(\theta,z_1,z_2)$ or the
$C_\ell(z_1,z_2)$'s contain the full three dimensional 2-point clustering information, exactly like the power spectrum.
However, the distinct advantage of these quantities w.r.t the commonly 
used power spectrum is that we need not assume any
background cosmology in order  to infer them from the observations.

In our approximation $\De(\bn,z)$ contains two terms which contribute  to $\xi$
with their correlators.
We indicate them as $D$ for the density term and $z$ for the
redshift space distortion,
\bea
\xi^{\rm gal}(\theta,z_1,z_2) &=&  b_1b_2\xi_{DD}(\theta,z_1,z_2) +b_1\xi_{Dz}(\theta,z_1,z_2) \nonumber \\
 && +b_2\xi_{zD}(\theta,z_1,z_2) +\xi_{zz}(\theta,z_1,z_2)\,,   \label{e:xidec}
\eea
where $b_i\equiv b(z_i)$ is the bias which only multiplies the density but not the velocity. 
We assume negligible velocity bias.
We now compute these terms within linear cosmological perturbation theory.
For this we denote by $P(k)$ the density-density power spectrum today and by
$G(z)$ the growth rate, such that $G(z=0)=1$ and $P(k,z) = G^2(z)P(k)$.
This product ansatz is justified under the assumption of vanishing sound speed (pure matter fluctuations) such
that all modes evolve with the same growth function.
It is certainly sufficient in standard $\La$CDM at $z\lsim 10$,
but might have to be revised for different dark energy models or if massive
neutrinos are taken into account~\cite{Eisenstein:1999a}.
If the neutrino masses are of the order of the measured mass differences,
they are not very relevant for structure formation, but if their mass is
larger, they lead to damping of small scale power.
More precisely $P(k)$ and $G(z)$ are defined by
\bea\label{e:Pdef}
\langle D(\bk, t_0)D^*(\bk',t_0)\rangle &=& (2\pi)^3\de(\bk-\bk')P(k)\\
  D(\bk, t(z)) &=& G(z) D(\bk, t_0)\,.
\eea
The comoving distance $r(z)$ from the observer to redshift $z$ in a fixed cosmology
is given by
\be\label{e:dz}
r(z) =\int_0^z\frac{dz'}{H(z')} \,.
\ee
This distance is related to the angular diameter distance via $D_A(z) = r(z)/(1+z)$ and to the luminosity distance
via $D_L(z) = r(z)(1+z)$.

Assuming vanishing spatial curvature, $K=0$, the law of cosines  implies that  the comoving distance $r(z_1,z_2,\theta)$ 
between two points at arbitrary redshifts $z_1$ and $z_2$ seen under an angle $\theta$ is
\be\label{e:rr}
r(z_1,z_2,\theta) = \sqrt{r(z_1)^2 + r(z_2)^2 - 2r(z_1)r(z_2)\cos\theta} \,.
\ee
With this we can express the angular correlation function $\xi(\theta,z_1,z_2)$
in terms of the spatial correlation function $\xi(r,z_1,z_2)$,
\be\label{xirtheta}
\xi(\theta,z_1,z_2) = \xi\left(r(z_1,z_2,\theta),z_1,z_2\right) .
\ee
This is useful since $\xi$ is simply the Fourier transform of the power spectrum. For $\xi_{DD}$ we now obtain
\be\label{e:xiDD}
\xi_{DD}(\theta,z_1,z_2) = G_1G_2\!\int_0^\infty\hspace{-2mm} \frac{dk}{2\pi^2} k^2P(k)j_0(kr(z_1,z_2,\theta)) \,,
\ee
where $G_j=G(z_j)$.
In a real galaxy survey, we observe a number of galaxies within a certain
redshift bin and within a solid angle. We assume that on sufficiently large scales,
the galaxy number fluctuations are related to the underlying density fluctuation 
by a scale-independent bias factor $b(z)$ and the velocities are unbiased.
Using the continuity equation, the redshift space distortion term can also be 
expressed in terms of the power spectrum $P$ and we find
\bea
&&\xi_{\rm gal}(\bn_1,\bn_2,z_1,z_2) = \langle \De_{\rm gal}(\bn_1,z_1) \De^*_{\rm gal}(\bn_2,z_2) \rangle = \nonumber \\
&&\quad b_1G_1b_2G_2 \int \frac{d^3k}{(2\pi)^3} P(k) e^{i\bk\cdot\br} \nonumber \\
&&\times \left[ 1+\frac{\beta_1}{3}+\frac{2\beta_1}{3}P_2(\bn_1\cdot\bnk)-\frac{i\al_1\HH_1\beta_1}{r_1k}P_1(\bn_1\cdot\bnk)  \right] \nonumber \\
&&\times \left[ 1+\frac{\beta_2}{3}+\frac{2\beta_2}{3}P_2(\bn_2\cdot\bnk)+\frac{i\al_2\HH_2\beta_2}{r_2k}P_1(\bn_2\cdot\bnk)  \right] \;, \nonumber \\
\label{eq:corrDD}
\eea
where $\bnk=\bk/k$ and $P_j(x)$ is the Legendre polynomial of degree $j$.
We also use $\br=\br_1-\br_2$, $\al_i\equiv\al(z_i)$, $\HH_i\equiv\HH(z_i)$,
and $\beta_i\equiv f(z_i)/b(z_i)$, where $f(z)\equiv d\log G(a(z))/d\log a(z)$,
$a(z)=1/(1+z)$ being the scale factor. The second term in the brackets of equation (\ref{eq:corrDD}) 
contributes to the well known plane-parallel approximation of RSD
\cite{Kaiser1987,Hamilton1992}.
From now on we neglect contributions to the function $\alpha(z)$
other than the RSD-related term $2/\HH$, so that the third and fourth terms
in the brackets just correct the plane-parallel approximation and
render the formula valid for arbitrary separation angles.

\begin{figure}[ht]
\begin{center}
\includegraphics[width=\columnwidth]{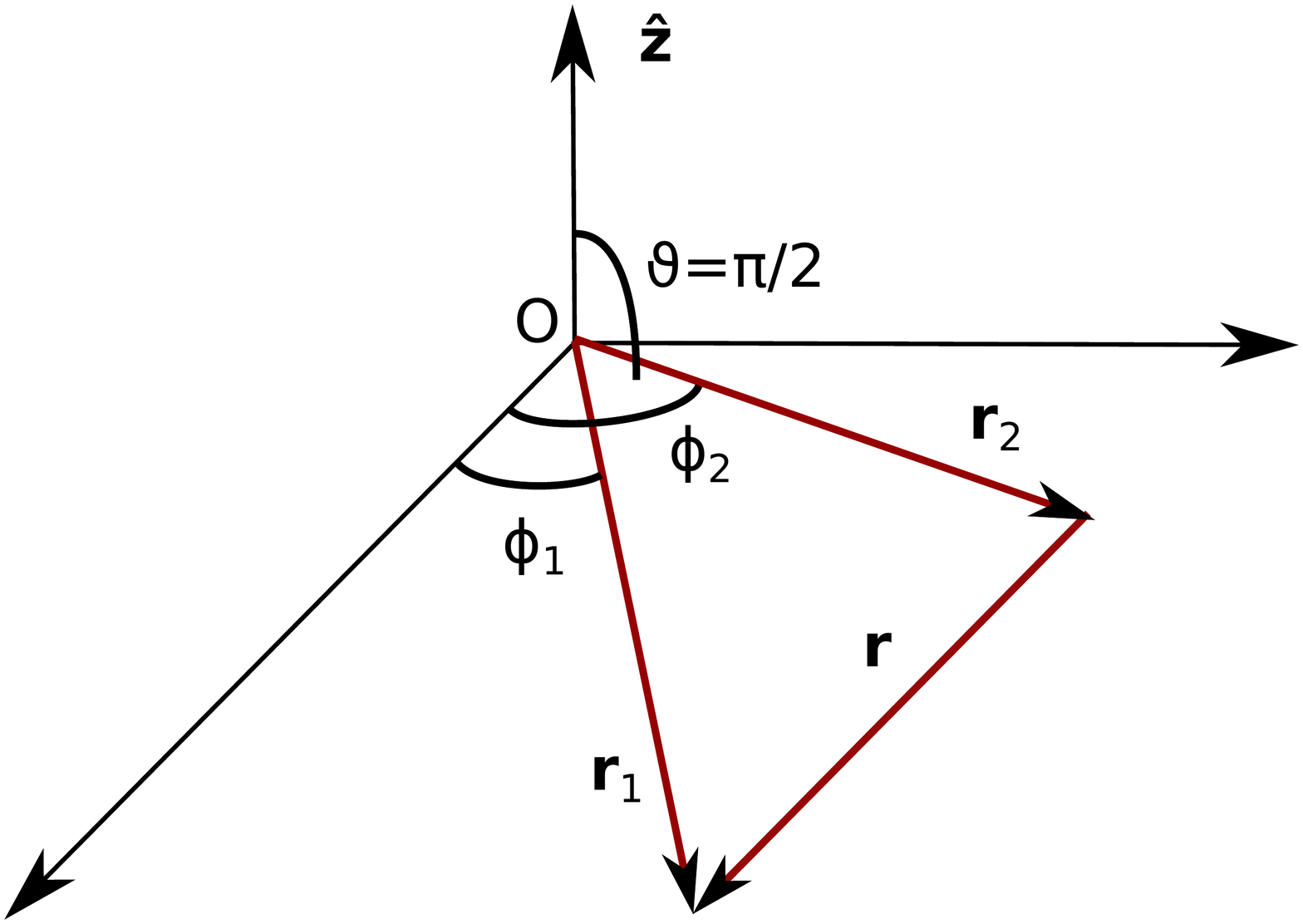}
\end{center}
\caption{Coordinate system. The triangle formed by the observer $O$ and the
   two galaxies is orthogonal to the direction of observation $\hat{z}$.
   We assume $\phi_1\leq\phi_2$.}
\label{fig:3polar_CoordSyst}
\end{figure}

We define the functions
\be
\zeta_\ell^m(r) \equiv \int \frac{dk}{2\pi^2} k^m j_\ell(kr) P(k) \;,
\label{e2:xi_lm}
\ee
where $j_\ell(x)$ denotes the spherical Bessel function of index $\ell$.
The results are obtained in terms of such integrals with $0\leq\ell\leq 4$
and $0\leq m\leq 2$ multiplied by functions of the redshift and of $\theta$.
This can be expressed in (relatively) compact form using an expansion
in tripolar spherical harmonics~\cite{Szapudi:2004gh,Papai:2008bd}. We present the
details in Appendix \ref{sec:tripolar}.
Here we just explain the basic idea: it is useful to introduce coordinates where
the triangle consisting of the observer and the two positions given by $S_1 =(\bn_1,z_1)$
and  $S_2 =(\bn_2,z_2)$ lies in the plane $\vth =\pi/2$ (the $x$-$y$ plane) 
and the direction $\overrightarrow{(S_2,S_1)}$ is parallel to the  $x$-direction. 
We also define $S_1$ as the position with the smaller azimuthal angle,
hence $\phi_1\le \phi_2$. This is depicted in Fig.~\ref{fig:3polar_CoordSyst}.

Then  $\theta\equiv\phi_2-\phi_1$, and 
the angles $\phi_1$, $\phi_2$ can be expressed in terms of the observables $z_1,~z_2$ and $\theta$ by
\bea
&&\phi_2 = \sin^{-1} \sbra{\frac{r_1}{r}\sin\theta} \;, \nonumber \\
&&\phi_1 = \phi_2 - \theta = \sin^{-1} \sbra{\frac{r_2}{r}\sin\theta} \;.
\label{eq:phi1phi2}
\eea
Here $r_j=r(z_j)$ and $r$ is the distance given in Eq.~(\ref{e:rr}).
 The constraint $\phi_1\leq\phi_2$ can be inverted using the symmetry
$\xi_{\rm gal}(\bn_2,\bn_1,\bn,r) = \xi_{\rm gal}(-\bn_1,-\bn_2,\bn,r)$.
Using this coordinate system, the tripolar expansion presented in the Appendix only 
contains a small number of terms and we end up with an expression of the form
\bea
&&\xi_{\rm gal}(z_1,,z_2,\theta) = b(z_1)G(z_1)b(z_2)G(z_2) \nonumber \\
&&\qquad\times \sum_{n_1,n_2=0,1,2} \left[a_{n_1n_2} \cos(n_1\phi_1)\cos(n_2\phi_2) \right. \nonumber \\
&&\phantom{\times \sum_{n_1,n_2=0,1,2}} \left. + b_{n_1n_2} \sin(n_1\phi_1)\sin(n_2\phi_2) \right] \;. 
\label{eq:corrDD_3d}
\eea
In a true observation we cannot measure the number of galaxies at a precise redshift, but we actually work with
redshift bins around the
fiducial redshifts $\bar{z}_1$, $\bar{z}_2$, where the probability distribution around the redshift 
$\bar z$ is given by a window function, typically a Gaussian
\be
W(z,\bar{z}) \propto \exp\sbra{-\frac{1}{2}\bra{\frac{z-\bar{z}}{\sigma_z}}^2} \;,
\ee
with standard deviation $\sigma_z$, centered around the mean
redshift $\bar{z}$ and normalized to 1.
The observable angular correlation function is then given by
\bea
\xi^{\rm g}(\theta,\bar{z}_1,\bar{z}_2) &=& \int dz_1 W(z_1,\bar{z}_1)b({z}_1)G(z_1)\times \nonumber \\
&&\int dz_2 W(z_2,\bar{z}_2)b({z}_2)G(z_2)\times \nonumber \\
&&\Xi(\phi_1(z_1,z_2,\theta),\phi_2(z_1,z_2,\theta),r(z_1,z_2,\theta)) \;, \nonumber \\
\label{e2:w_3p}
\eea
where we have introduced
\bea
\Xi(\phi_1,\phi_2,r) &\equiv& \sum_{n_1,n_2=0,1,2} \left[a_{n_1n_2} \cos(n_1\phi_1)\cos(n_2\phi_2) \right. \nonumber \\
&&\left. + b_{n_1n_2} \sin(n_1\phi_1)\sin(n_2\phi_2) \right] \;.
\label{eq:Xi_def}
\eea
The difference between $\xi_{\rm gal}$ and $\xi^{\rm g}$ is that the latter has been smeared over redshifts with a
window function of widths $\si_z$. 
The  non-vanishing coefficients  $a_{n_1,n_2}$ and $b_{n_1,n_2}$  
are given in Eqs.~(\ref{eq:coord_coeff}) and 
(\ref{eq:coord_coeff_alpha}) in terms of the integrals $\zeta^m_\ell$. 

In order to understand the signal in clustering analysis of
galaxy surveys, we must also estimate the error bars in the
measurements.
Since we are interested on scales where non-linear effects do not play a
prominent role, we can model the expected errors following \cite{Crocce:2011}.
The covariance in the measurements of $\xi^{\rm g}(\theta,\bar{z}_1,\bar{z}_2)$,
i.e.,
${\rm Cov}_{\xi\xi'}\equiv\braket{\tilde{\xi}^{\rm g}(\theta,\bar{z}_1,\bar{z}_2) \tilde{\xi}^{\rm g}(\theta',\bar{z}'_1,\bar{z}'_2)}$, where $\tilde{\xi}^{\rm g}$ denotes the
estimator used for $\xi^{\rm g}$,
can be related to the angular power spectrum, defined in equation (\ref{e:cl}),
as
\bea
\label{e:cov}
{\rm Cov}_{\xi\xi'} &=& \frac{2}{f_{\rm sky}}\sum_{\ell\ge 0} \frac{2\ell +1}{(4\pi)^2}
  P_{\ell}(\cos\theta) P_{\ell}(\cos\theta') \bra{C_{\ell}^{\rm g}+1/\bar{N}}^2 \nonumber \\
&=&\sum_{\ell\ge 0} \frac{2\ell +1}{(4\pi)^2}
  P_{\ell}(\cos\theta) P_{\ell}(\cos\theta') {\rm Cov_{\ell,\ell}(z,z')}\,.
\eea
This includes the effects of sampling variance, shot-noise ($\bar{N}$ is the
number of objects per steradian), partial sky coverage ($f_{\rm sky}$ is the
observed fraction of the sky), photometric redshift uncertainties and RSD.
We have introduced the power spectra smeared over a redshift bin.
\bea
C_{\ell}^{\rm g}(\bar{z}_1,\bar{z}_2) &\equiv& \int dz_1 W(z_1,\bar{z}_1)b({z}_1) \times\nonumber \\
&&\int dz_2 W(z_2,\bar{z}_2)b({z}_2) C_{\ell}(z_1,z_2)
\eea
This angular power spectrum has been discussed in \cite{Bonvin:2011bg} and,
according to their notation, we are approximating
$C_{\ell} = C^{DD}_{\ell} + C^{Dz}_{\ell} + C^{zz}_{\ell}$, where the three
terms refer to the correlations between density-density, density-RSD,
and RSD-RSD, respectively.
We emphasize that if we write the covariance matrix in terms of reconstructed
distances instead of observable angles and redshifts, we would not obtain
a consistent error estimation. 
Also, the covariance matrix of an ideal experiment is 
diagonal in $\ell$ but not in $z$ and $z'$.
This complicates the analysis as compared, e.g. to CMB spectra.

We shall consider the detailed computation of this covariance matrix for
a given experiment in future
work, whereas here we focus on the correlation function.

From a numerical point of view, equation (\ref{e2:w_3p}) may look quite
cumbersome, since it involves a sum of three dimensional integrals,
two of which over redshift and one over wavelengths,
see eq.~(\ref{e2:xi_lm}).
However, the functions $\zeta_\ell^m(x)$ depend only on the cosmology so,
given the cosmological parameters, they can be
calculated once and then stored.
Therefore, eq.~(\ref{e2:w_3p}) requires only a two-dimensional
integral of the window functions over redshifts, which can be performed rapidly.
Then, eqs.~(\ref{eq:coord_coeff}), which are the leading coefficients at
BAO scales (see Fig. \ref{fig:RSD_residuals} below), require the evaluation of the only three
functions $\zeta_0^2(x)$ (which is also called the `real space' correlation function), 
$\zeta_2^2(x)$ and $\zeta_4^2(x)$.
With this, eq.~(\ref{e2:w_3p}) demands the same
numerical effort as the plane-parallel case, where the evaluation of three
functions similar to $\zeta_\ell^m(x)$'s is needed~\cite{Hamilton1992}.
However, eq.~(\ref{e2:w_3p}) takes fully into account wide-angle effects that are
neglected in the plane-parallel approximation.

To evaluate eq.~(\ref{e2:w_3p}), we compute the power spectrum 
 using the {\sc Camb} code \cite{Lewis:1999bs}.
The linear result is used to obtain the $\zeta_\ell^m(r)$'s,
equation (\ref{e2:xi_lm}), on large scales $r\gtrsim 150\ {\rm Mpc}/h$,
while {\sc Halofit} \cite{Halofit} is used on smaller scales to take
 into account corrections for non-linear evolution approximately,
as in \cite{Challinor2005PhRvD} .

In the examples presented in this paper we use the cosmological parameters 
$\Om_m=0.24$, $\Om_b=0.04$, $\Om_{\Lambda}=0.76$,
$h=0.73$, primordial amplitude and spectral index equal to
$A_s=1.99\cdot 10^{-9}$, $n_s=0.96$, respectively,
and a constant bias $b=1$.

\begin{figure}[ht]
\begin{center}
\includegraphics[width=\columnwidth]{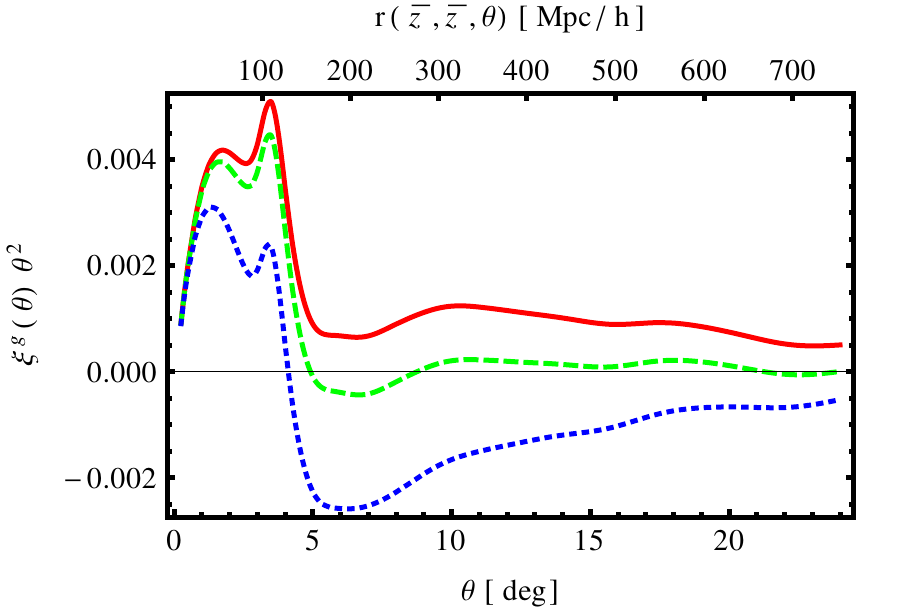}
\end{center}
\caption{The angular correlation function (\ref{e2:w_3p}) at $\bar z =0.7$
multiplied with $\theta^2$ (in degrees) is shown.
The different curves correspond, from top to bottom, to the result in redshift
space discussed above (top solid line), the plane-parallel approximation (middle dashed line) and the real space
correlation function $\xi^{\rm g}_{\rm DD}$ (bottom, short-dashed line), respectively.}
\label{fig:w_trans_WideAngle}
\end{figure}
In Figure \ref{fig:w_trans_WideAngle} the angular correlation function
integrated over redshift bins, Eq.~(\ref{e2:w_3p}), is shown.
A Gaussian selection function compatible with photometric surveys
$\sigma_z/(1+z)=0.03$ is employed, and the bins are centered around the mean
redshifts $\bar{z}_1=\bar{z}_2=0.7$.
The top axis has been evaluated using equation (\ref{e:rr}).
For better visibility of the BAO peak we plot $\theta^2 \xi^{\rm g}(\theta)$.
The results in redshift space differ significantly from the real space correlation function.
The plane-parallel approximation \cite[equation (10)]{Szapudi:2004gh},
does not give an accurate description for angles
$\theta\gtrsim 1^{\rm o}$, compared to the wide-angle RSD result obtained
with equation (\ref{eq:corrDD_3d}).

Another interesting observation is that once we correctly include redshift space distortions, 
the correlation function no longer passes through zero. The enhanced clustering 
observed in redshift space, has a correlation function which is everywhere positive 
at fixed observed redshift. For the pure real space 
correlation function this behavior is not possible due to the integral constraint,
$$ \int r^2\xi_{DD}(r)dr =0 \,.$$
As the correlation function is no longer isotropic when accounting for RSD,
this constraint no longer holds for the observed $\xi^{\rm g}$.
The large-scale anti-correlation in
real space is  due to the initial conditions set up at the end
of inflation and evolution under gravity.
The transition  of a Harrison-Zeldovich from $P(k) \propto k$ on large scales to $P (k) \propto k^{-3}$
spectrum  is due to the fact that density perturbations
can grow only in the matter era. For modes that enter the horizon during the radiation
dominated era (see, e.g., \cite{Durrer:2008aa}), this delays the growth roughly until matter-radiation 
equality, so that $k^3P(k)$ is roughly constant on these scales.
The scale of the horizon at matter-radiation equality is imprinted as the 
 location  of this turnover in the power spectrum~\cite{Eisenstein:1999} and
equivalently,  as the location of the zero-crossing of the real space correlation function~\cite{Prada:2011}.

However, along the transverse direction, RSD suppress anti-correlations at
large scales.
In fact RSD introduce additional positive correlations between velocity and the density 
contrast at fixed observed redshift. These additional positive correlations are larger than the
negative density correlations.
This is shown in Figure \ref{fig:w_trans_WideAngle}, where a relatively large
anti-correlation at $\theta_{\rm min}\approx 6^{\rm o}$ in the real space correlation function 
$\xi_{\rm DD}$,
$\abs{\xi^{\rm g}_{\rm DD}(\theta_{\rm min})\,\theta_{\rm min}^2}\approx  2.5\times 10^{-3}\,{\rm deg}^2$,
is lifted to a small positive correlation in redshift-space
$\abs{\xi^{\rm g}(\theta_{\rm min})\,\theta_{\rm min}^2}\approx  7.5\times 10^{-4}\,{\rm deg}^2$.

\begin{figure}[ht]
\begin{center}
\includegraphics[width=\columnwidth]{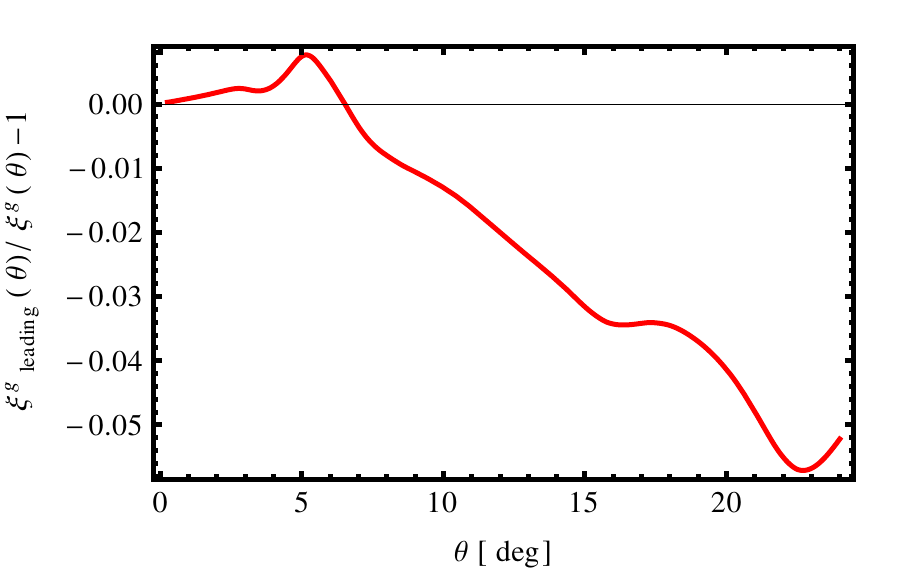}
\end{center}
\caption{Residuals of the redshift space correlation functions calculated
as in Figure \ref{fig:w_trans_WideAngle} neglecting (i.e., considering only
leading RSD contribution) and taking into account the sub-leading
coefficients in equation (\ref{eq:coord_coeff_alpha}) for $\al(z)=2/\HH(z)$.}
\label{fig:RSD_residuals}
\end{figure}
The curve in Figure \ref{fig:RSD_residuals} suggests that the full RSD result,
with $\al(z)=2/\HH(z)$,
differs by up to $6\%$ from the leading contribution (i.e., with $\al(z)=0$)
only for large angles. However, on angular scales up to $\theta\approx 7^{\rm o}$, 
which is well above the BAO's peak, the difference is less than 1\%.
This justifies to neglect $\al(z)$ in the next section, where we are interested in 
the BAO scale.

%-------------------------------------------------------------------------------
\section{ The Alcock-Paczy\'nski test}\label{s:ap}
One of the deepest enigmas of modern cosmology is the problem of dark energy: The energy density in the present 
Universe seems to be dominated by a cosmological constant $\La$ providing about 70\% to the expansion 
rate of the Universe. For a review of the dark energy problem see, e.g.~\cite{Durrer:2007re}. So far, most indications
for dark energy come from measurements of the distance redshift relation~\cite{Durrer:2011gq}, and therefore rely
on the Friedmann equation and on relation~(\ref{e:dz}). This relation can be tested if we can measure both, 
$r(z)$ and  $H(z)$ independently.  Alcock and Paczy\'nski have proposed such a measurement as follows: imagine a spherical object of  comoving size $L$ in the sky at redshift $z$. The redshift difference of its front and back is
then given by $\De z_L(z) = LH(z)$ and we see it under an angle
$\theta_L(z) = \frac{L}{(1+z)D_A(z)}$.
Knowing $L$ we can in principle determine both $(1+z)D_A(z)$ and $H(z)$ by measuring 
$\De z_L$ and $\theta_L$. But even without any knowledge of $L$ we can infer the product
\be \label{e:ap}
F(z)\equiv (1+z)H(z)D_A(z) = \frac{\De z_L(z)}{\theta_L(z)} \equiv F^{AP}(z) \,.
\ee
Combining this with a measurement of the luminosity distance $D_L(z)=(1+z)^2D_A(z)$,
e.g. from supernova type 1a data~\cite{Suzuki:2011hu}, we can break the degeneracy between 
$H(z)$ and $D_A(z)$. This allows us to test the  relation
$$  D_A(z) = \frac{1}{z+1}\int_0^z\frac{dz'}{H(z')} \quad \mbox{or}\quad F(z) = \int_0^zdz'\frac{H(z)}{H(z')}\,$$
which must be valid, if the geometry of our Universe is close to a Friedmann-Lema\^\i tre metric.

Assuming an incorrect cosmology for the determination of the galaxy power spectrum
causes geometric redshift-distortions, in addition to
the dynamical redshift distortions due to peculiar velocities of the galaxies
discussed here.
Hence, the observed clustering signal can be used to constrain the underlying
cosmology (Alcock–Paczy\'nski test \cite{Alcock:1979mp}).
This is also why it is important to understand the dynamical RSD.
As suggested also, e.g., in \cite{Gaztanaga2009a}, we want to propose the baryon acoustic
oscillations (BAO) scale $r_{\rm BAO}$ in the galaxy correlation function
to provide the scale $L$.

When considering physical distances, a cosmology must be assumed and, e.g.,
iterative methods have been proposed to converge to the correct
cosmology \cite{Guzzo:2008}.
The inferred galaxy clustering in a different cosmological model
can also be obtained from the fiducial one by a rescaling of the
transverse and parallel separations \cite{Percival:2010}.
However here, instead of using these approximate procedures,
we shall not assume a cosmology to determine  the power spectrum
but we work only with the directly observable redshift dependent 
angular correlation function $\xi^{\rm g}(\theta,z_1,z_2)$.

In this work we discuss how to constrain the quantity $F(z)=(1+z)H(z)D_A(z)$.
The degeneracy between $D_A(z)$ and $H(z)$ can then be lifted e.g. when 
comparing this with the direction averaged BAO scale~\cite{Eisenstein:2005}, 
$D_A^2(z)/H(z)$, or with SN Ia distances.

For an introduction to BAO's see~\cite{Eisenstein:1997ik,Montanari:2011nz}.
It has been suggested before to use the  BAO's for an Alcock-Paczy\'nski test, e.g.,
in~\cite{Simpson:2009zj,Hawken:2011nd}.
The BAO ring (observed in~\cite{Okumura:2008}) has been studied in \cite{Hu:2003,Matsubara:2004fr},
and \cite{Padmanabhan:2008} analyses the anisotropies affecting the BAO's
without assuming a particular form of RSD.
Experimental realizations of the Alcock-Paczy\'nski test analysing the clustering
anisotropies are reported in~\cite{Blake:2011ep,Kazin2012,Reid:2012sw}.
In \cite{Gaztanaga2009a} the possibility to observe
BAO's along the line-of-sight (LOS), allowing a direct measurement of $H(z)$ has been studied.
Although the BAO detection along the LOS in the LRG catalog of the
SDSS DR6 survey used in this reference has been called into question
\cite{Kazin2010a}, in \cite{Tian2011} an enhancement along the LOS has been
observed in the SDSS DR7 MGS which is compatible with a BAO feature.
Furthermore, \cite{Benitez2009b} argued that a direct measurement of $H(z)$
thanks to the BAO peak along the LOS might be possible with a photometric
survey such that $\sigma_z/(1+z)=0.003$.

As an example of what can be achieved with the angular correlation function, we
now study the transversal and longitudinal BAO's, using the angular and redshift correlation.
function obtained in the previous section.

%-------------------------------------------------------------------------------
\subsection{Transverse correlation function.}
The transverse correlation function
is obtained from equation (\ref{e2:w_3p}) by considering two bins at the same
mean redshift $\bar{z}$ and varying their angular separation
$\theta=\phi_2-\phi_1$:
\bea
\xi^{\rm g}(\theta,\bar{z},\bar{z}) &=& \int dz_1 W(z_1,\bar{z})b(z_1)G(z_1) \nonumber \\
&&\times\int dz_2 W(z_2,\bar{z})b(z_2)G(z_2) \nonumber \\
&&\times \Xi(\phi_1,\phi_2,r(z_1,z_2,\theta)) \;,
\label{eq:w_trans}
\eea
where $\phi_1$ and $\phi_2$ are given in equation (\ref{eq:phi1phi2}), and
$\Xi$ in equation (\ref{eq:Xi_def}).
As said, numerically it is convenient to first evaluate the $\zeta_\ell^m(r)$'s,
equation (\ref{e2:xi_lm}).
Since these functions only depend on the cosmology (through the power spectrum
$P(k,z=0)$), they can be stored and used efficiently for a fixed cosmology.

We expect that RSD increases the amplitude of the transversal correlation
function with respect to the real space result.
For example, a spherical distribution of galaxies in real space is squashed
along the line of sight by peculiar velocities.
Since galaxies are collapsing toward the center of the distribution,
those between its center of and the observer receive an
additional red-shift from RSD, while those
beyond the center are blueshifted.
When observing the transversal correlation function, this leads to
increase of the amplitude, since over-dense regions are
enhanced.

\begin{figure}[hT]
\begin{center}
\includegraphics[width=\columnwidth]{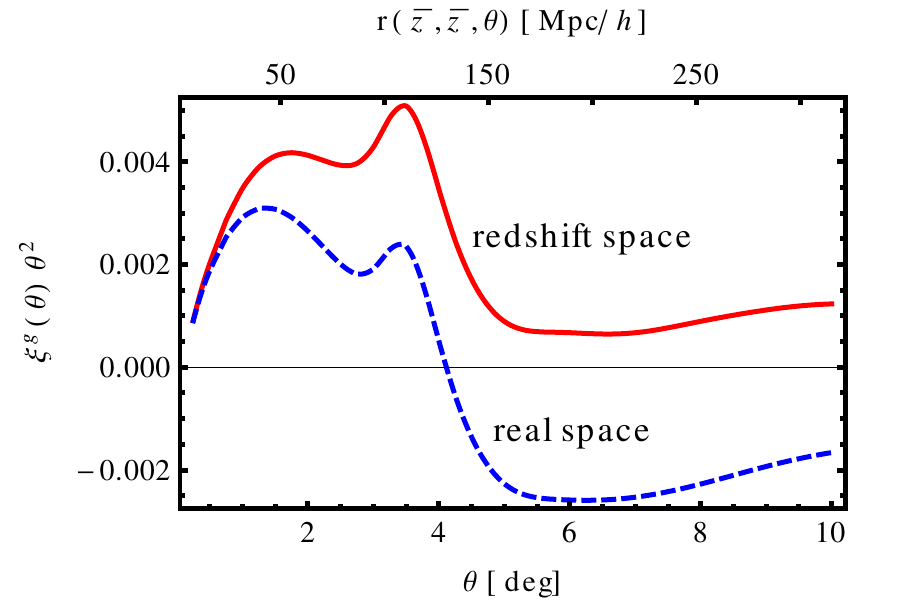}
\end{center}
\caption{The transversal correlation function.
Two redshift bins with mean redshift $\bar{z}_1=\bar{z}_2=0.7$ are considered.
The window function is a Gaussian with $\sigma_z/(1+z)=0.03$.
Non-linearities are approximated with {\sc Halofit}.
RSD is relevant for the shape of the correlation function, and,
in particular, it shifts the location of the BAO peak.}
\label{fig:w_trans}
\end{figure}
Figure \ref{fig:w_trans} confirms that the increase in clustering amplitude
due to RSD along the transverse direction is a relevant effect.
The transverse correlation function is obtained using a Gaussian radial
window function with $\sigma_z/(1+z)=0.03$, compatible with the
photometric requirements of, e.g., Euclid \cite{EuclidRB,EuclidWeb}.
As a reference, equation (\ref{e:rr}) has been used on the top axis to give
the comoving scale corresponding to a certain angle.
The BAO peak at $\theta\simeq 3.45^{\rm o}$ corresponds
to about $110\ {\rm Mpc}/h$.
The real space curve is obtained by neglecting all the coefficients depending
on $\b(z)$ in equation (\ref{eq:w_trans}).
The difference in the positions of BAO peak is of order $1\%$ between real
space and redshift space.
\begin{figure}[ht]
\begin{center}
\includegraphics[width=\columnwidth]{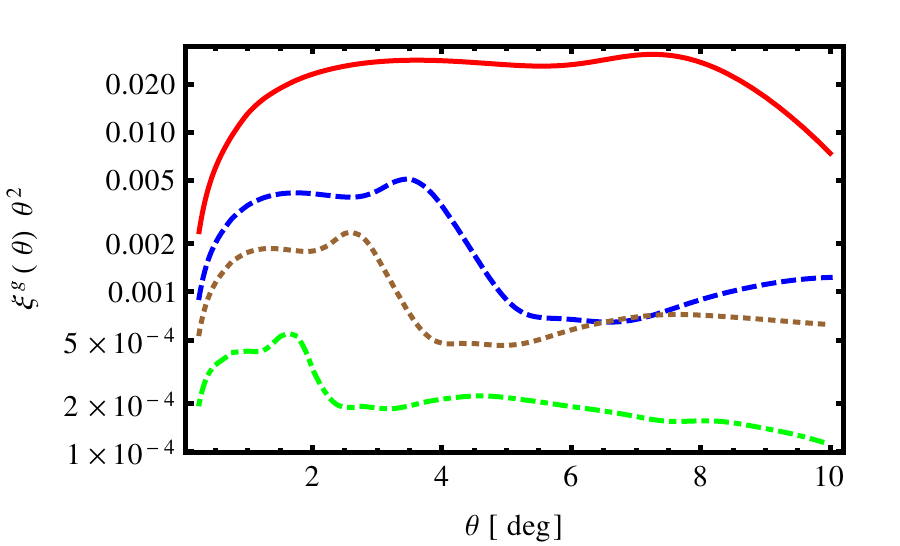}
\end{center}
\caption{Transversal correlation functions.
The curves correspond, from top to bottom, to mean redshifts
$\bar{z}=0.3,\ 0.7,\ 1,\ 2$, respectively.
A Gaussian radial window function with $\sigma_z/(1+\bar z)=0.03$ is used.
Non-linearities are approximately considered with {\sc Halofit}.}
\label{fig:w_trans_dtheta}
\end{figure}
In Figure \ref{fig:w_trans_dtheta} we show the transverse correlation function
for different redshifts.
The BAO peak is located at
$\theta=7.2^{\rm o},\ 3.5^{\rm o},\ 2.6^{\rm o},\ 1.6^{\rm o}$
for $\bar{z}=0.3,\ 0.7,\ 1,\ 2$, respectively, which consistently corresponds to about
110 Mpc$/h$ in terms of the comoving separation, equation (\ref{e:rr}).
As the redshift $\bar{z}$ increases, the angle under which we observe
the comoving BAO scale $r_{\rm BAO}$ decreases.
This angle is given by $\theta_L(z) = \frac{L}{(1+z)D_A(z)} = L/r(z)$,
where $r(z)$, given in equation (\ref{e:dz}), monotonically increases with redshift
(note that this is not the case for $D_A(z)$ that shows a turnover at 
$z\approx 2$).
The peak of the curve corresponding to $\bar{z}=0.3$ is smeared out
due to the logarithmic scale, which is used because of the large
change in the amplitude of the correlation functions.

%-------------------------------------------------------------------------------
\subsection{Longitudinal correlation function}
The longitudinal correlation function is obtained by fixing
$\theta=\phi_2-\phi_1=0$:
\bea
\xi(0,\bar{z}_1,\bar{z}_2) &=& \int dz_1 W(z_1,\bar{z}_1)b(z_1)G(z_1)  \nonumber \\
&&\times\int dz_2 W(z_2,\bar{z}_2)b(z_2)G(z_2) \nonumber \\
&&\times \Xi(0,0,r(z_1,z_2,0)) \;,
\label{eq:w_long}
\eea
where
\be
\Xi(0,0,r) = \sum_{n_1,n_2=0,1,2} a_{n_1n_2} \;.
\ee

\begin{figure}[ht]
\begin{center}
\includegraphics[width=\columnwidth]{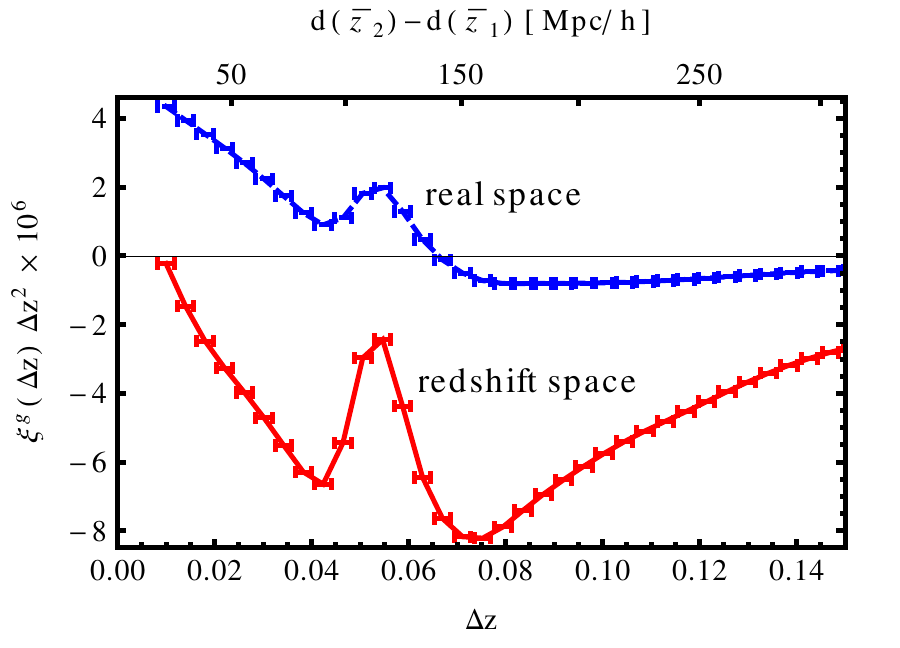}
\end{center}
\caption{The longitudinal correlation function multiplied by $\De z^2$ is shown.
The redshift $\bar{z}_1=0.7$ is fixed and we vary $\De z=\bar{z}_2-\bar{z}_1$.
A Gaussian radial window function with $\sigma_z/(1+\bar z)=0.001$ is used.
Error bars along the abscissas correspond to $\pm\sigma_z$.
Non-linearities are approximately considered with {\sc Halofit}.}
\label{fig:w_long}
\end{figure}

We expect  different effects from RSD on the
longitudinal correlation function than on the transversal one.
Along the line of sight the over-densities are enhanced,
but they are also squashed to smaller scales.
Instead, along the transversal direction, the enhanced over-dense regions
still extend over the same scales as in real space.
The squashing in redshift space implies that, along the line of sight,
the amplitude of the correlation function on large (small) scales is
reduced (enhanced) with respect to the real space result.
Figure \ref{fig:w_long} shows that, for our cosmological parameters,
the correlation function becomes even entirely negative on scales
larger than $\De z\simeq 0.01$, corresponding to about $20\ {\rm Mpc}/h$
at $\bar z = 0.7$.
Note that on smaller scales non-linear effects becomes important, and to
compare with observation a treatment of the non-linear the 'fingers-of-god' effect
is necessary.

Fig.~\ref{fig:w_long} is obtained from equation (\ref{eq:w_long}).
We set $\bar{z}_1=\bar z-\De z/2$ and $\bar{z}_2=\bar z+\De z/2$ and plot 
$\xi^{\rm g}(0,z_1,z_2)$ as function of $\De z$ for $\bar z=0.7$.
A Gaussian radial window function with $\sigma_z/(1+z)=0.001$,
compatible with spectroscopic requirements of Euclid \cite{EuclidRB,EuclidWeb},
is employed.
The functions $\zeta_\ell^m(r)$ are calculated with the specifications given 
in Fig.~\ref{fig:w_trans}, and the cosmological parameters are also the same.
As a reference, on the top axis equation (\ref{e:dz}) has been used to show
the comoving galaxy separation corresponding to a certain redshift separation.
The BAO peak at $\De z\simeq 0.055$ corresponds to about $115\ {\rm Mpc}/h$.
The BAO scale is only very mildly affected by RSD. The difference of the peak position
with and without RSD as shown Fig.~\ref{fig:w_long} is less than $1\%$.
For better visibility of the BAO peak we plot $\De z^2 \xi^{\rm g}(\De z)$.
It is clear
from the figure that a redshift resolution which is significantly
better that $\si_z \sim 0.03(1+z)$ is needed to resolve the BAO feature.
Therefore photometric redshift with the accuracy proposed for Euclid would 
not suffice. However, if photometric redshift can be improved to a level of
$\si_z \sim 0.003(1+z)$ as proposed in~\cite{Benitez2009b},
one might be able to see  also the longitudinal BAO's with
photometric redshifts.
\begin{figure}[ht]
\begin{center}
\includegraphics[width=\columnwidth]{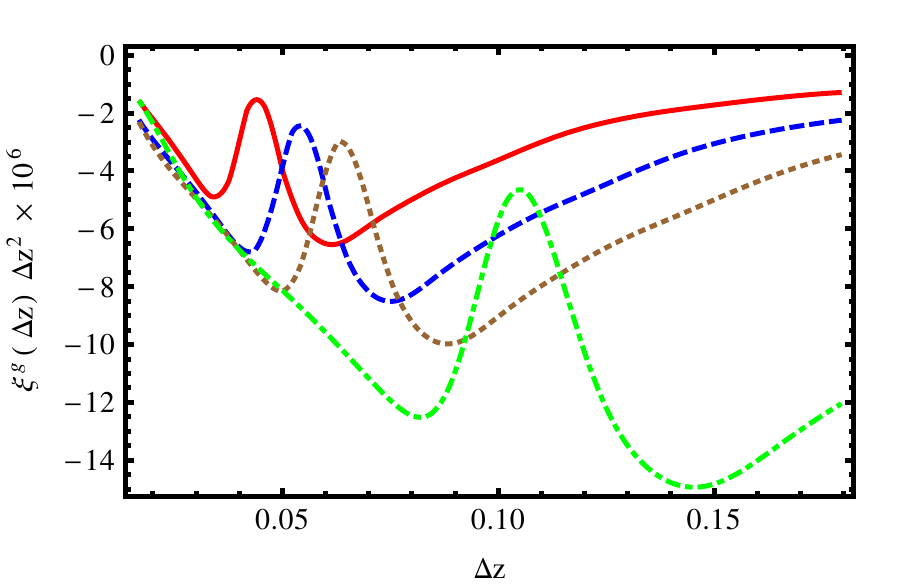}
\end{center}
\caption{Longitudinal correlation functions with different mean redshifts.
The curves, with the BAO peak from the left to right,
correspond to initial redshifts $\bar{z}_1=0.3,\ 0.7,\ 1,\ 2$, respectively.
A Gaussian radial window function with $\sigma_z/(1+z)=0.001$ is used.
Non-linearities are approximately considered with {\sc Halofit}.}
\label{fig:w_long_dz}
\end{figure}
In Figure \ref{fig:w_long_dz} we compare the longitudinal correlation
functions at different redshifts.
The BAO's peak is located at
$\De z=0.044,\ 0.054,\ 0.064,\ 0.106$
for $\bar{z}_1=0.3,\ 0.7,\ 1,\ 2$, respectively, which correspond to about
115 Mpc$/h$ in terms of the comoving separation $r(\bar{z}_2)-r(\bar{z}_1)$.
The BAO peak shifts to higher $\Delta z  = \bar{z}_2-\bar{z}_1$ as the
redshift $\bar{z}$, at which the transversal correlation function is computed,
increases according to equation (\ref{e:dz}).

It is also interesting that the longitudinal correlation function is fully negative for 
$\De z\gsim 0.01$ and when multiplied by $(\De z)^2$ it becomes nearly redshift 
independent before the raise to the acoustic peak.
As said before, the sign is due to the fact that in longitudinal direction clustering is severely squashed
due to redshift space distortion as it is visible also in direction dependent plots which
can be found e.g. in Ref.~\cite[figure 3]{Reid:2012sw} or~\cite[figure 1]{Kazin2010a}.
On scales larger than the small redshift difference to which clustering is squashed, 
correlations along the line of sight are negative.

\begin{figure}[ht]
\begin{center}
\includegraphics[width=\columnwidth]{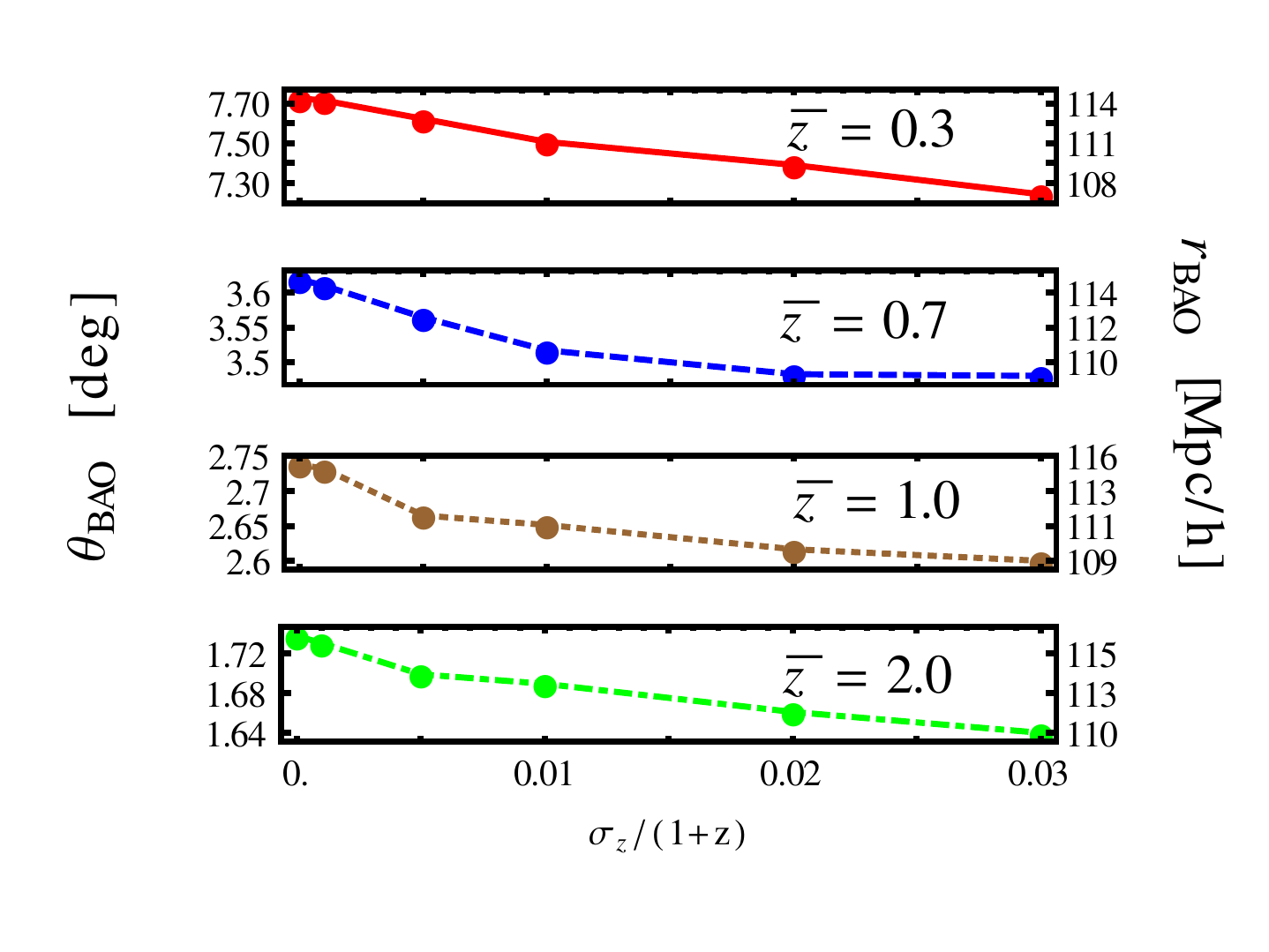}
\end{center}
\caption{The angle corresponding to the BAO's peak in the transverse correlation
function is shown for different
redshifts $\bar{z}$ in function of the
parameter $\si_z/(1+z)$, which determines the width of the radial selection
function.
As a reference, on the right it is also indicated the comoving distance
according to the $\La$CDM model.}
\label{fig:Theta_window}
\end{figure}

\begin{figure}[ht]
\begin{center}
\includegraphics[width=\columnwidth]{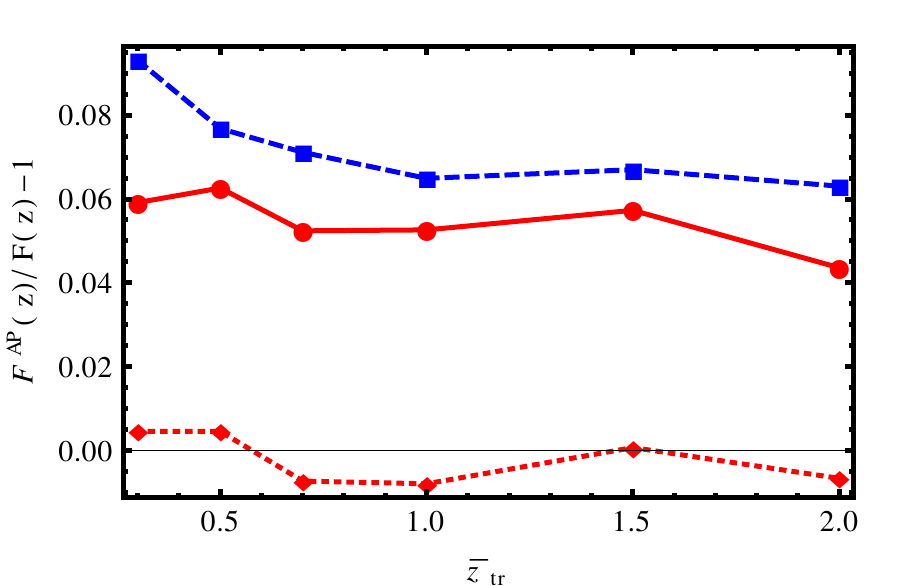}
\includegraphics[width=\columnwidth]{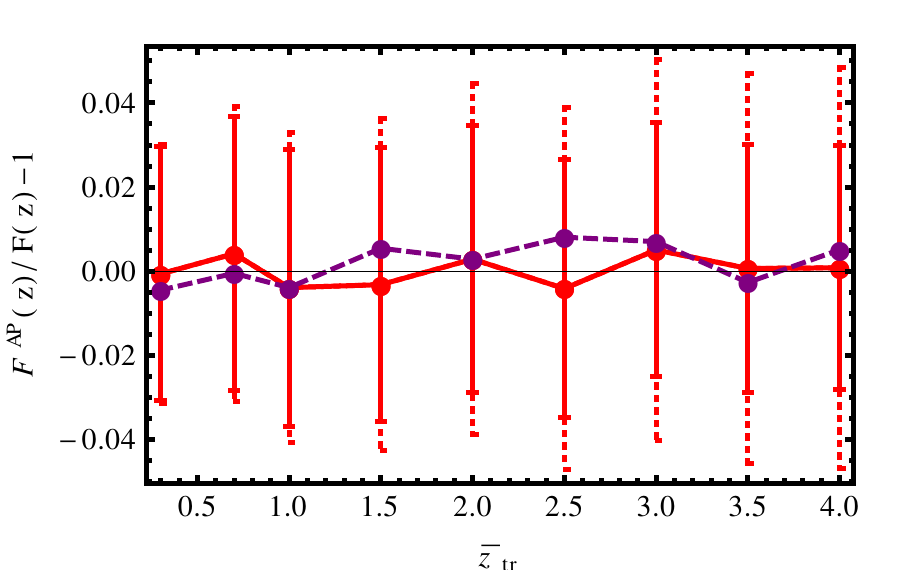}
\end{center}
\caption{\emph{Top panel:} $z$-space (solid) and real-space (long-dashed) AP
consistency test for $\sigma_z/(1+z)=0.03$. The short-dashed line takes into account
the correction $F^{\rm AP}_{\rm corr}$ with $\ga=1/8$.
\emph{Bottom panel:}
Result without radial window function (solid line). Solid error bars consider
angular resolution $\de\theta=0.02^{\rm o}$, and dotted error bars
$\de\theta=0.05^{\rm o}$.
The dashed line consider a linear $P(k)$.
For illustrational purposes, the analysis is extrapolated up to
$\bar{z}_{\rm tr}=4$.}
\label{fig:Fap_F}
\end{figure}

The BAO scale inferred from the longitudinal correlation function is $L\sim 115$ Mpc$/h$. This is nearly 5\% larger 
than the value expected from the transversal correlation
function, $L_\perp \sim 110\ {\rm Mpc}/h$. This systematic difference is mainly due to the 
redshift selection function. If the redshift bin is relatively wide, the transverse BAO scale is reduced
 by projecting $L$ transversal to the line of sight to $L_\perp<L$. This effect 
 can be considerable.
$L_\perp$ converges to the longitudinal scale $L$ for 
$\si_z\ra 0$, see Fig.~\ref{fig:Theta_window}.
To take into account that at finite redshift resolution the BAO peak is not fully transversal,
we correct the corresponding scale $L$ by adding a small longitudinal component $L_{\parallel} \sim \de zH^{-1}(z)$.
\bea\label{e:corL}
L &=& \sqrt{\left(\frac{\de z}{H(z)}\right)^2 + L_\perp^2} = \De z_L/H(z)\,,\\
 L_\perp &=& \theta_L(z)D_A(1+z)\,.
\eea
This takes into account an averaged longitudinal contribution.
Solving this expression for $F(z)$ we obtain
\bea
F(z) &=& \frac{\De z_L(z)}{\theta_L(z)}\sqrt{1 -\left(\frac{\de z}{\De z_L}\right)^2} \nonumber\\
&\simeq&  \frac{\De z_L(z)}{\theta_L(z)} \left[1 -\frac{1}{2}\left(\frac{\de z}{\De z_L}\right)^2 \right]  \nonumber\\   
&=& F^{AP}(z)\left[1 -\frac{\ga}{2}\left(\frac{\si_z}{\De z_L}\right)^2\right]    \nonumber\\   
&\equiv& F^{AP}_{\rm cor}(z) \,. \label{e:corF}
\eea
Here we have set $\de z =\sqrt{\ga}\si_z$. Naively one might expect $\ga\sim 1$, however, we have found that the correction is significantly smaller and requires only $\ga=1/8$, see top panel of Fig.~\ref{fig:Fap_F} . We suggest that this comes from the fact that  longitudinally separated galaxies are actually anti-correlated on this scale and therefore the contributions
from galaxies with significant longitudinal separation are suppressed in the positive transversal correlation function.
However, we have not found a satisfactory derivation of the pre-factor $\ga=1/8$.
 
In the top panel of Fig.~\ref{fig:Fap_F} we compare this redshift-corrected Alcock-Paczy\'nksi function, 
$F^{AP}_{\rm cor}(z)$ with $\ga=1/8$ and the uncorrected function $F^{AP}(z)$.
Also the real-space result is shown, which is even larger.
Clearly, the difference from the input 
function $F(z)$ is significantly reduced by this redshift-correction.
From eq.~(\ref{e:corL}) it is clear, the smaller $\sigma_z$ the smaller the difference
between the longitudinal and the transversal BAO scale. 
To remove the effect of the redshift bin and check for other systematics, we plot
in Fig.~\ref{fig:Fap_F} (lower panel) the real space
result for $F^{AP}(z)/F(z)$ with redshift bins of zero width,
i.e. using $\xi^{\rm gal}(\theta,z_1,z_2)$ without RSD terms
to determine the BAO  peak position.
We have also found that it is crucial to choose symmetric longitudinal redshift differences,
$\xi^{\rm gal}(0,\bar{z}-\De z/2,\bar{z}+\De z/2)$,
where $\bar{z}$ is the redshift at which the transverse correlation
function is computed, and we vary $\De z$.
Other choices, e.g.
$\xi^{\rm gal}(0,\bar{z},\bar{z}+\De z)$,
do not lead to a consistent result because the longitudinal and transversal
correlation functions compute the BAO's at slightly different redshifts
$\bar{z}$.
Furthermore, we have found that the peak position is significantly improved 
by using good angular resolution. We compare the result of the linear power spectrum
and the one using {\sc halofit} to take into account non-linear effects.
Finally we add formal error bars using a redshift resolution of
$\de\De z =0.001(1+z)$
from the Euclid spectral survey and an angular resolution of $\de\theta =0.05^o$ and $0.02^o$.
Clearly, for $z\ge 1$ the angular resolution significantly affects the errors. 
 Assuming these formal errors, and allowing for several redshift bins, 
a large survey such as Euclid should enable us to measure
$F(z)$ with a few percent accuracy.

%-------------------------------------------------------------------------------
\section{Conclusions}\label{s:con}
In this work we have introduced the directly measurable 
redshift dependent angular correlation function. We argue that this function contains the full
three dimensional information about galaxy clustering. The standard
spatial correlation function requires the assumption of a cosmology and a
determination of cosmological parameters with it can therefore 
at best be considered as a consistency test.
We have also found that the shapes of the longitudinal and transverse correlation 
function are not only very different from each other, but also very different from the 
real space correlation function. While the integral of the latter has to vanish, so that
$\xi(r)<0$ for large $r$, we have found that $\xi(\theta,z,z)>0~\forall~\theta$ and
$\xi(0,z-\De z/2,z+\De z/2)<0~\forall~ \De z>0.01$.
 
We used the redshift dependent angular correlation function to
determine the transverse and the longitudinal acoustic peak positions as function
of the redshift. We have proposed to use these functions to perform an 
Alcock-Paczy\'nski test. We have fully taken into account redshift space distortions. 
Even though these modify significantly the shape of the correlation function, they 
are not very important for the position of the acoustic peak. 

We have found that
photometric redshift determinations with $\si_z/(z+1) \simeq 0.03$ are sufficient
to determine the BAO peak in the transverse correlation function, but in order to
cleanly determine the position of the longitudinal peak,  spectroscopic precision with
$\si_z/(z+1) \simeq 0.001$ are needed. Furthermore, we have seen that the peak
position of the transverse correlation function slightly depends on the window function 
for the redshift determination. To achieve an accuracy better than about 6\% for 
the transversal BAO scale, one has the employ the correction suggested in this work.
This, together with the correctly symmetrized longitudinal correlation
function allows to obtain an estimation of
$F(z)=(1+z)H(z)D_A(z)$ which is accurate to about 2\%, when performing the Alcock-Paczy\'nski test.

\acknowledgments{We thank Chris Clarkson, Enea Di Dio, Roy Maartens and
Rom\'an Scoccimarro for discussions.
We also aknowledge Camille Bonvin for sharing with us her codes, and the
anonymous Referee for pointing out the lack of discussion about the
covariance matrix in a previous draft of the paper.
The GNU Scientific Library (GSL) has been used for the numerical implementation.
This work is supported by the Swiss National Science Foundation.}
%_______________________________________________________________________________
\appendix
%-------------------------------------------------------------------------------
\section{Wide-angle RSD using the tripolar expansion} \label{sec:tripolar}
In this appendix we derive expressions for the parameters $a_{n_1 n_2}$ and $b_{n_1 n_2}$ used
in the calculation of the correlation function via Eq.~(\ref{eq:Xi_def}).

We use the coordinates introduced in section~\ref{s:gen} and shown in Fig.~\ref{fig:3polar_CoordSyst}.
We can characterize the correlation function, equation (\ref{eq:corrDD}),
by the two directions $\bn_1$, $\bn_2$ from the observer and the
comoving distance between the two galaxies $\br=\br_1-\br_2 =r\bn$.
As pointed out in \cite{Hamilton:1996px,Szalay:1998cc}, the correlation
function in redshift space depends on the triangle formed by the observer and
the two galaxies.
This triangle is invariant under rotation about the observer and hence can
be described by one size and two shape parameters.
For the former we use the separation of the galaxies $r$, while for the shapes
we choose the angles $\phi_1$, $\phi_2$, e.g., between $\bn_1$, $\bn_2$
and $\bn$, respectively, see Fig.~\ref{fig:3polar_CoordSyst}.
We suppose that $\phi_1\leq\phi_2$ and the distances are $r_1=r(z_1)$ and $r_2=r(z_2)$.
For fixed redshifts $z_1$ and $z_2$, the distance $r$ is constrained by
$|r_1-r_2|\le r \le r_1+r_2$. 
We can evaluate the galaxy correlation function as given in Eq.~(\ref{eq:corrDD}),
\bea
\xi_{\rm gal}(\bn_1,\bn_2,r) \equiv
  &&\quad b_1G_1b_2G_2 \int \frac{d^3k}{(2\pi)^3} P(k) e^{i\bk\cdot\br} \nonumber \\
&&\hspace{-2.3cm}\times \left[ 1+\frac{\beta_1}{3}+\frac{2\beta_1}{3}P_2(\bn_1\cdot\bnk)-\frac{i\al_1\HH_1\beta_1}{r_1k}P_1(\bn_1\cdot\bnk)  \right] \nonumber \\
&&\hspace{-2.3cm}\times \left[ 1+\frac{\beta_2}{3}+\frac{2\beta_2}{3}P_2(\bn_2\cdot\bnk)+\frac{i\al_2\HH_2\beta_2}{r_2k}P_1(\bn_2\cdot\bnk)  \right] .  \nonumber \\ &&   \label{eq:xiDeDe_def}
\eea
The most direct way would be to replace the occurrences of $\bnk$ in the
integrand of equation (\ref{eq:xiDeDe_def}) in terms of the gradient $\nabla_\br$.
This is indeed straightforward in the plane-parallel limit, 
because Fourier modes are eigenfunctions of the
plane-parallel distortion operator,
that corresponds to the case in which the terms of equation (\ref{eq:corrDD})
multiplied by Legendre polynomials are negligible,
and can be written in terms of the
inverse Laplacian $\nabla_r^{-2}$ \cite{Hamilton:1996px}.
However, it is non-trivial  to obtain an expression for
$\xi_{\rm gal}$ beyond the plane-parallel limit proceeding in this way.
In fact, the computation would lead to a large number of
terms that, although straightforward, are cumbersome to evaluate numerically.
Despite the non-locality of the power spectrum including the wide-angle RSD,
the corresponding correlation function has still been calculated
in spherical coordinates and expressed in a closed form, e.g., in
\cite{Szalay:1998cc} by the introduction of a convenient spherical tensor and
in the approximation of small redshifts $z\ll 1$.
This work has been generalized by \cite{Matsubara:1999du, Matsubara:2004fr}
to the case of
arbitrary galaxy redshifts $z_1$, $z_2$ also for non-flat cosmologies
(the $\al$-term there is a selection function).
Even with these results, however, the symmetry of the problem, namely the
dependence of $\xi_{\rm gal}$ on the rotationally invariant triangle
determined by $\left\{\bn_1,\bn_2,\bn\right\}$, suggests that a more natural
way to proceed is to expand the expression in tripolar spherical harmonics.
In particular, we use the functions defined in~\cite{Szapudi:2004gh,Papai:2008bd}
\bea
S_{\ell_1\ell_2\ell}(\bn_1,\bn_2,\bn) &=& \sum_{m_1,m_2,m} \tj{\ell_1}{\ell_2}{\ell}{m_1}{m_2}{m} \nonumber \\
&\times& C_{\ell_1m_1}(\bn_1)C_{\ell_2m_2}(\bn_2)C_{\ell m}(\bn) \;, \nonumber \\
\label{eq:S_l1l2l}
\eea
where $C_{\ell m}(\bn)=\sqrt{\frac{4\pi}{2\ell+1}} Y_{\ell m}$ are conveniently normalized
spherical harmonics and
$(\begin{smallmatrix}
  \ell_1 & \ell_2 & \ell \\
  m_1 & m_2 & m
\end{smallmatrix})$
is the  Wigner 3-$j$ symbol.
These function form an complete orthogonal  basis for expanding
spherically symmetric functions depending on three vectors, which are invariant under global rotations.
In~\cite{Szapudi:2004gh,Papai:2008bd} it has been shown that, in the
approximation $z\ll 1$, this expansion leads to a more compact form for
$\xi_{\rm gal}$ than those obtained in previous works.
We generalize the work of \cite{Szapudi:2004gh,Papai:2008bd}
by allowing arbitrary galaxy redshifts $z_1$, $z_2$
and a generic function $\al(z)$.

After some algebra one finds that the only non-vanishing coefficients of the expansion
\bea
&&\xi_{\rm gal}(\bn_1,\bn_2,\bn,r) = b_1G_1b_2G_2 \nonumber \\
&&\quad \times\sum_{\ell_1\ell_2\ell} B^{\ell_1\ell_2\ell}(r,z_1,z_2)
   S_{\ell_1\ell_2\ell}(\bn_1,\bn_2,\bn) \;,
\label{eq:xi_tp}
\eea
are 
\bea
&&B^{000}(r,z_1,z_2) = \bra{1+\frac{1}{3}\b_1} \bra{1+\frac{1}{3}\b_2} \;\zeta_0^2(r) \;, \nonumber \\
&&B^{220}(r,z_1,z_2) = \frac{4}{9\sqrt{5}} \b_1\b_2 \;\zeta_0^2(r)\;, \nonumber \\
&&B^{222}(r,z_1,z_2) = \frac{4\sqrt{10}}{9\sqrt{7}} \b_1\b_2 \;\zeta_2^2(r)\;, \nonumber \\
&&B^{224}(r,z_1,z_2) = \frac{4\sqrt{2}}{\sqrt{35}} \b_1\b_2 \;\zeta_4^2(r)\;, \nonumber \\
&&B^{202}(r,z_1,z_2) = -\bra{\frac{2}{3}\b_1+\frac{2}{9}\b_1\b_2} \sqrt{5} \;\zeta_2^2(r)\;, \nonumber \\
&&B^{022}(r,z_1,z_2) = -\bra{\frac{2}{3}\b_2+\frac{2}{9}\b_1\b_2} \sqrt{5} \;\zeta_2^2(r)\;,
\label{eq:Bl1l2l}
\eea
which are independent of the angles (as they do not involve the $\al$-terms), and
\bea
&&B^{101}(r,z_1,z_2) = -\frac{\al_1\HH_1\sqrt{3}}{r_1} \bra{\b_1+\frac{\b_1\b_2}{3}} \;\zeta_1^1(r) \;, \nonumber \\
&&B^{011}(r,z_1,z_2) = \frac{\al_2\HH_2\sqrt{3}}{r_2} \bra{\b_2+\frac{\b_1\b_2}{3}} \;\zeta_1^1(r) \;, \nonumber \\
&&B^{121}(r,z_1,z_2) = \frac{\al_1\HH_1}{r_1} \frac{2\sqrt{2}}{\sqrt{15}} \b_1\b_2 \;\zeta_1^1(r) \;, \nonumber \\
&&B^{123}(r,z_1,z_2) = \frac{\al_1\HH_1}{r_1} \frac{2\sqrt{7}}{\sqrt{15}} \b_1\b_2 \;\zeta_3^1(r) \;, \nonumber \\
&&B^{211}(r,z_1,z_2) = -\frac{\al_2\HH_2}{r_2} \frac{2\sqrt{2}}{\sqrt{15}} \b_1\b_2 \;\zeta_1^1(r) \;, \nonumber \\
&&B^{213}(r,z_1,z_2) = -\frac{\al_2\HH_2}{r_2} \frac{2\sqrt{7}}{\sqrt{15}} \b_1\b_2 \;\zeta_3^1(r) \;, \nonumber \\
&&B^{110}(r,z_1,z_2) = -\frac{\al_1\HH_1\al_2\HH_2}{r_1r_2} \frac{\b_1\b_2}{\sqrt{3}} \;\zeta_0^0(r) \;, \nonumber \\
&&B^{112}(r,z_1,z_2) = -\frac{\al_1\HH_1\al_2\HH_2}{r_1r_2} \sqrt{\frac{10}{3}} \b_1\b_2 \;\zeta_2^0(r) \;. \nonumber \\
\label{eq:Bl1l2l_al}
\eea
If one consider $\b_1=\b_2=f$, i.e., neglecting the bias and setting $z_1=z_2$,
and $\al_1=\al_2=2/\HH$, this result agrees with \cite{Papai:2008bd}.
We also verified the consistency with \cite{Bertacca2012tp}.

To proceed further we must choose a system of coordinates.
As shown in \cite{Szapudi:2004gh}, however, once the coefficients of the
tripolar expansion have been calculated, it is trivial to write the correlation
function for a given coordinate system.
In fact, this step involves only simple algebra that, although tedious, can
easily be carried out using a computer algebra package.

We use the coordinate system a) discussed in \cite{Szapudi:2004gh} since it gives
the most compact form for $\xi_{\rm gal}$.
As shown in Figure \ref{fig:3polar_CoordSyst}, the direction
$\hat{\bf z}$ is orthogonal to the plane that contain the triangle formed by
the observer $O$ and the two galaxies.
Hence, all the vectors $\br_1$, $\br_2$, $\br$ have latitude $\vartheta=\pi/2$.
The galaxy separation vector $\br$ has zero longitude $\phi=0$, i.e., it is parallel
to the abscissas.

With this choice of coordinates we can evaluate the elements of the tripolar
basis, equation (\ref{eq:S_l1l2l}), which so far have been written in terms
of the re-normalized spherical harmonics $C_{\ell m}(\bn)$. Using the 
notation $\bn=\bbra{\vartheta,\phi}$,
 we have
$$
S_{\ell_1\ell_2\ell}(\bn_1,\bn_2,\bn) =
   S_{\ell_1\ell_2\ell}(\bbra{\pi/2,\phi_1},\bbra{\pi/2,\phi_2},\bbra{\pi/2,0}) \;.
$$
Evaluating these expressions, the summation over $\ell_1,\ell_2$ and $\ell$ in Eq.~(\ref{eq:xi_tp}), using
(\ref{eq:Bl1l2l}) and (\ref{eq:Bl1l2l_al}) for the non-vanishing functions $B^{\ell_1\ell_2\ell}$ leads to 
\bea
&&\xi_{\rm gal}(z_1,\phi_1,z_2,\phi_2,r) = b(z_1)G(z_1)b(z_2)G(z_2) \nonumber \\
&&\qquad\times \sum_{n_1,n_2=0,1,2} \left[a_{n_1n_2} \cos(n_1\phi_1)\cos(n_2\phi_2) \right. \nonumber \\
&&\qquad\phantom{\times \sum_{n_1,n_2=0,1,2}} \left. + b_{n_1n_2} \sin(n_1\phi_1)\sin(n_2\phi_2) \right] \;. \nonumber \\
\label{eA:corrDD_3d}
\eea
The coefficients $a_{n_1n_2}$ and $b_{n_1n_2}$ can be easily calculated with 
a numerical algebra package. We first write all the non-vanishing coefficients which do not involve $\al$-terms:
\bea
a_{00}&=&\bra{1+\frac{1}{3}\bra{\b_1+\b_2}+\frac{2}{15}\b_1\b_2} \zeta_0^2(r) \nonumber \\
&& - \bra{\frac{1}{6}\bra{\b_1+\b_2}+\frac{2}{21}\b_1\b_2} \zeta_2^2(r) \nonumber \\
&& + \frac{3}{140}\b_1\b_2 \;\zeta_4^2(r) \;, \nonumber \\
a_{20}&=& -\bra{\frac{1}{2}\b_1+\frac{3}{14}\b_1\b_2} \zeta_2^2(r) + \frac{1}{28}\b_1\b_2 \;\zeta_4^2(r) \;, \nonumber \\
a_{02}&=& -\bra{\frac{1}{2}\b_2+\frac{3}{14}\b_1\b_2} \zeta_2^2(r) + \frac{1}{28}\b_1\b_2 \;\zeta_4^2(r) \;, \nonumber \\
a_{22}&=& \frac{1}{15}\b_1\b_2 \;\zeta_0^2(r) - \frac{1}{21}\b_1\b_2 \;\zeta_2^2(r) + \frac{19}{140}\b_1\b_2 \;\zeta_4^2(r) \;, \nonumber \\
b_{22}&=& \frac{1}{15}\b_1\b_2 \;\zeta_0^2(r) - \frac{1}{21}\b_1\b_2 \;\zeta_2^2(r) - \frac{4}{35}\b_1\b_2 \;\zeta_4^2(r) \;, \nonumber \\
\label{eq:coord_coeff}
\eea
The coefficients that involve $\al$-terms 
 are:
\bea
a_{10}&=& \bra{\al_1\HH_1\b_1+\frac{2}{5}\al_1\HH_1\b_1\b_2}\frac{1}{r_1} \zeta_1^1(r) \nonumber \\
&& - \frac{\al_1\HH_1}{10}\frac{\b_1\b_2}{r_1} \zeta_3^1(r) \;, \nonumber \\
a_{01}&=& -\bra{\al_2\HH_2\b_2+\frac{2}{5}\al_2\HH_2\b_1\b_2}\frac{1}{r_2} \zeta_1^1(r) \nonumber \\
&& + \frac{\al_2\HH_2}{10}\frac{\b_1\b_2}{r_2} \zeta_3^1(r) \;, \nonumber \\
a_{11}&=& \frac{\al_1\HH_1\al_2\HH_2}{3}\frac{\b_1\b_2}{r_1r_2} \zeta_0^0(r) \nonumber \\
&& - \frac{2\al_1\HH_1\al_2\HH_2}{3}\frac{\b_1\b_2}{r_1r_2} \zeta_2^0(r) \;, \nonumber \\
a_{21}&=& -\frac{\al_2\HH_2}{5}\frac{\b_1\b_2}{r_2} \zeta_1^1(r) + \frac{3\al_2\HH_2}{10}\frac{\b_1\b_2}{r_2} \zeta_3^1(r) \;, \nonumber \\
a_{12}&=& \frac{\al_1\HH_1}{5}\frac{\b_1\b_2}{r_1} \zeta_1^1(r) - \frac{3\al_1\HH_1}{10}\frac{\b_1\b_2}{r_1} \zeta_3^1(r) \;, \nonumber \\
b_{11}&=& \frac{\al_1\HH_1\al_2\HH_2}{3}\frac{\b_1\b_2}{r_1r_2} \zeta_0^0(r) \nonumber \\
&&+ \frac{\al_1\HH_1\al_2\HH_2}{3}\frac{\b_1\b_2}{r_1r_2} \zeta_2^0(r) \;, \nonumber \\
b_{21}&=& -\frac{\al_2\HH_2}{5}\frac{\b_1\b_2}{r_2} \zeta_1^1(r) - \frac{\al_2\HH_2}{5}\frac{\b_1\b_2}{r_2} \zeta_3^1(r) \;, \nonumber \\
b_{12}&=& \frac{\al_1\HH_1}{5}\frac{\b_1\b_2}{r_1} \zeta_1^1(r) + \frac{\al_1\HH_1}{5}\frac{\b_1\b_2}{r_1} \zeta_3^1(r) \;. \nonumber \\
\label{eq:coord_coeff_alpha}
\eea
Setting $\b_1=\b_2=f$ and $\al_1=\al_2=2/\HH$, we recover the results
of~\cite{Papai:2008bd}.
We recall that we are assuming $\phi_1\leq\phi_2$
(Figure \ref{fig:3polar_CoordSyst}), this constraint
can be inverted using the symmetry
$\xi_{\rm gal}(\bn_2,\bn_1,\bn,r) = \xi_{\rm gal}(-\bn_1,-\bn_2,\bn,r)$
evident from equation (\ref{eq:corrDD}).

\bibliography{AP-refs}
\bibliographystyle{h-physrev4}
\end{document}